# Cooper pair ring model

E.F. Talantsev[1,2]

[1]M.N. Mikheev Institute of Metal Physics, Ural Branch, Russian Academy of Sciences, 18, S. Kovalevskoy St., Ekaterinburg, 620108, Russia

[2]NANOTECH Centre, Ural Federal University, 19 Mira St., Ekaterinburg, 620002, Russia

E-mail: evgeny.talantsev@imp.uran.ru

*Abstract*

The superconducting state starts to collapse when the externally applied magnetic field exceeds the Meissner-Ochsenfeld critical field, $B_{c,MO}$, which in type-I superconductors is the thermodynamic critical field, $B_c = \frac{\phi_0}{2 \cdot \sqrt{2} \cdot \pi} \cdot \frac{1}{\lambda \cdot \xi}$, where λ is the London penetration depth, ξ is the coherence length, and $\phi_0$ is the flux quantum, while in type-II superconductors this field is the lower critical field, $B_{c1} = \frac{\phi_0}{4 \cdot \pi} \cdot \frac{ln\left(\frac{\lambda}{\xi}\right) + 0.5}{\lambda^2}$. Here we show that both critical fields can be described by the universal equation of $B_{c,MO} = \mu_0 \cdot n \cdot \mu_B \cdot ln\left(1 + \sqrt{2} \cdot \frac{\lambda}{\xi}\right)$, where $\mu_0$ is the magnetic permeability of free space, *n* is the Cooper pair density, and $\mu_B$ is the Bohr magneton. As a result, the Meissner-Ochsenfeld field can be defined as the field at which each Cooper pair exhibits the diamagnetic moment of one Bohr magneton with a multiplicative pre-factor of $ln\left(1 + \sqrt{2} \cdot \frac{\lambda}{\xi}\right)$. In the two-dimensional case this implies that the Cooper pair center of mass is spatially confined within a ring with inner radius $\xi$ and outer radius of $\xi + \sqrt{2} \cdot \lambda$. This means that the superconducting transition is associated not only with the charge carrier pairing, but that the pairs exhibit a new topological state with genus 1.

# Cooper pair ring model

## 1. Introduction

Walther Meissner and Robert Ochsenfeld discovered that an external magnetic field, *B*, is expelled from superconducting tin and lead [1]. Since then this effect is considered to be one of the most fundamental effects in superconductivity [2-10], which has been utilized in several superconducting technologies [11-16].

One of the key questions is how to quantitatively describe the maximum flux density at which the superconducting state starts to collapse. In type-I superconductors, for which the Ginzburg-Landau parameter $\kappa(T) = \frac{\lambda(T)}{\xi(T)} < \frac{1}{\sqrt{2}}$ (where $\lambda(T)$ is the London penetration depth and $\xi(T)$ is the coherence length), this field is given by the so-called thermodynamic critical field:

$$B_c(T) = \frac{\phi_0}{2\cdot\sqrt{2}\cdot\pi} \cdot \frac{1}{\lambda(T)\cdot\xi(T)} \tag{1}$$

where $\phi_0 = \frac{h}{2\cdot e} \approx 2.07 \cdot 10^{-15}$ Wb is the flux quantum, *h* is the Planck constant, and *e* is the electron charge. In type-II superconductors, for which the Ginzburg-Landau parameter $\kappa(T) \geq \frac{1}{\sqrt{2}}$, the field is called the lower critical field, $B_{c1}(T)$, for which Brandt proposed the expression [16]:

$$B_{c1}(T) = \frac{\phi_0}{4\cdot\pi} \cdot \frac{\ln(\kappa(T)) + \alpha(\kappa(T))}{\lambda^2(T)} \tag{2}$$

where

$$\alpha(\kappa(T)) = 0.49693 + e^{\left(-0.41477 - 0.775\cdot\ln(\kappa(T)) - 0.1303\cdot\left(\ln(\kappa(T))\right)^2\right)}. \tag{3}$$

More details about the lower critical field can be found elsewhere [17-19].

An empirical equation combining both Eq. 1 and Eq. 2 and which describes the maximum magnetic flux density, i.e. Meissner-Ochsenfeld critical field, $B_{c,MO}$, which the material can expel, has been proposed recently [20]:

$$B_{c,MO}(T) = \frac{\phi_0}{4\cdot\pi} \cdot \frac{\ln(1+\sqrt{2}\cdot\kappa(T))}{\lambda^2(T)} \tag{4}$$

Here we report the physical background to Eq. 4.

## 2. Bohr magnetons density in a superconductor

First we show that both Eq. 1 and Eq. 2 do not require for their definition the concept of magnetic flux quantum, $\phi_0$. Eq. 1 can be re-written in the form:

$$B_c(T) = \frac{h}{4\cdot\sqrt{2}\cdot\pi\cdot e} \cdot \frac{\kappa(T)}{\lambda^2(T)} = \mu_0 \cdot \left(\frac{\hbar\cdot e}{2\cdot m_e}\right) \cdot n(T) \cdot \sqrt{2} \cdot \kappa(T) = \mu_0 \cdot \mu_B \cdot n(T) \cdot \sqrt{2} \cdot \kappa(T) \tag{5}$$

where $\mu_B = \frac{\hbar\cdot e}{2\cdot m_e}$ is the Bohr magneton, and the bulk density of Cooper pairs in the material, $n$, is given by:

$$n(T) = \frac{1}{2} \cdot \frac{1}{\mu_0\cdot e^2} \cdot \frac{m_e}{\lambda^2(T)} \tag{6}$$

Similarly, Eq. 2 can be re-written in the form:

$$B_{c1}(T) = \frac{h}{8\cdot\pi\cdot e} \cdot \frac{\ln(\kappa(T))+\alpha(\kappa(T))}{\lambda^2(T)} = \mu_0 \cdot \mu_B \cdot n(T) \cdot \left(\ln(\kappa(T)) + \alpha(\kappa(T))\right), \tag{7}$$

and consequently the empirical Eq. 4 which describes both Type-I and Type-II superconductors can be re-written in the form:

$$B_{c,MO}(T) = \mu_0 \cdot \mu_B \cdot n(T) \cdot \ln\left(1 + \sqrt{2} \cdot \kappa(T)\right) \tag{8}$$

Eq. 8 can be interpreted to say that maximum diamagnetic response in the superconductor is achieved when each Cooper pair in the material exhibits a magnetic momentum of one Bohr magneton with a logarithmic multiplicative pre-factor of $\ln\left(1 + \sqrt{2} \cdot \kappa(T)\right)$. It should be noted that the Bohr magneton has the same value for single and double charges:

$$\mu_B = \frac{\hbar\cdot e}{2\cdot m_e} = \frac{\hbar\cdot(2\cdot e)}{2\cdot(2\cdot m_e)}. \tag{9}$$

This means that the magnetic moment of the Cooper pair:

$$\mu_{Cooper\ pair}(T) = \mu_B \cdot \ln\left(1 + \sqrt{2} \cdot \frac{\lambda(T)}{\xi(T)}\right) = \mu_B \cdot \ln\left(\frac{\xi(T)+\sqrt{2}\cdot\lambda(T)}{\xi(T)}\right) \tag{10}$$

requires a different physical interpretation, which we report for the two-dimensional (2D) case in the next Section.

### 3. Ring magneton

Consider a model, where a thin disk with a large concentric hole lies in the *xy* plane, centered on the origin (Fig. 1). The disk has inner radius *a* and outer radius *b*.

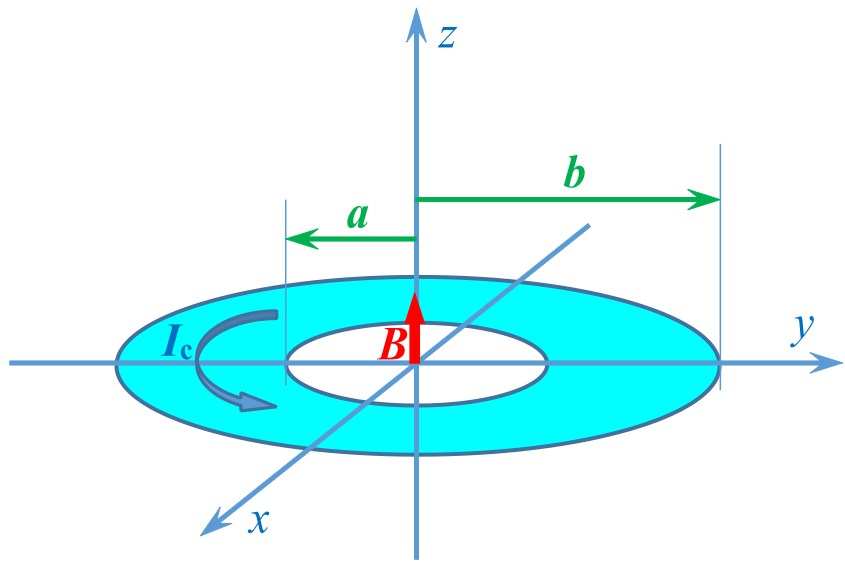

**Figure 1.** Schematic diagram of the model. The vector of magnetic flux density in the center of the disk, ***B***, is shown.

The disk carries a uniformly distributed surface current, $I_c$, by which we understand that the charge is distributed uniformly within the disc and the tangential charge velocity, $v_0$, is constant within the disc (Fig. 2,a). This means that each elemental concentric coil of radius *r* and width *dr* is carrying the same current, *dI*, independent of *r* (this is because the electric current is defined as the rate of electric charge flowing through the cross-section of the conductor (which in our 2D model is *dr*)). It should be noted that this current is created by the movement of the central mass of a Cooper pair.

In our model (Fig. 2,a), each thin disk with radius $r$ and width $dr$ creates a magnetic flux density, $dB$, in the center of the disc:

$$dB = \frac{\mu_0}{2} \cdot \frac{I_c \cdot dr}{b-a} \cdot \frac{1}{r} \tag{11}$$

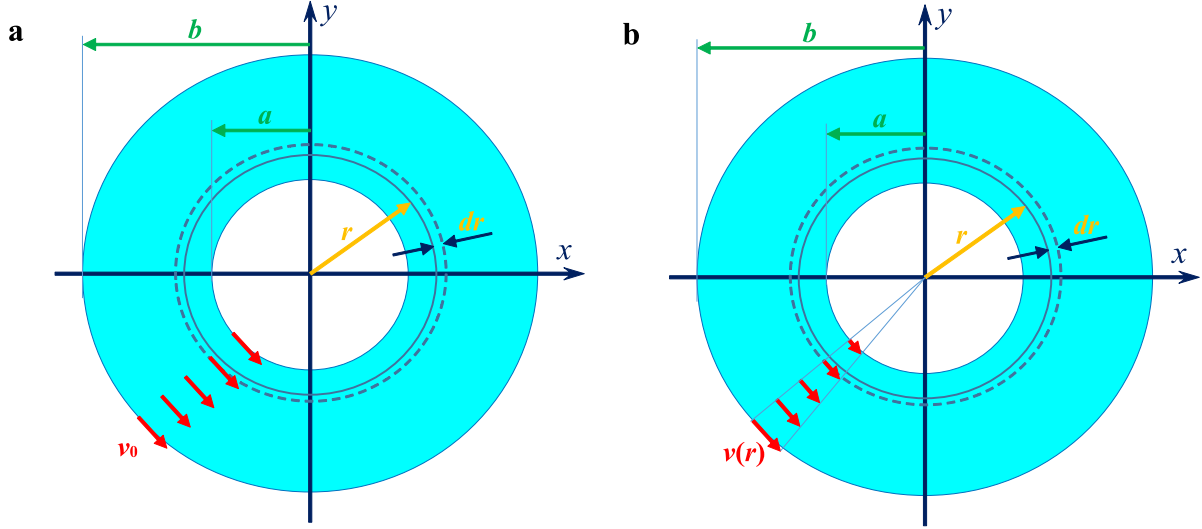

**Figure 2.** Plan view of the current-carrying disks. Red arrows indicate the charge velocity. **a** – proposed model; **b** – solid magneton model.

Integration of Eq. 11 gives the total magnetic flux density in the center of the disk:

$$\int_a^b dB = \frac{\mu_0}{2} \cdot \frac{I_c}{b-a} \cdot \int_a^b \frac{dr}{r} = \frac{\mu_0}{2} \cdot \frac{I_c}{b-a} \cdot \ln\left(\frac{b}{a}\right) \tag{12}$$

By comparing Eq. 12 with Eq. 9 it is natural to propose that:

$$\begin{cases} a = \xi \\ b = \xi + \sqrt{2} \cdot \lambda \end{cases} \tag{13}$$

In this case, Eq. 12 can be written in the form:

$$B_{center} = \frac{\mu_0}{2} \cdot \frac{I_c}{\sqrt{2} \cdot \lambda} \cdot \ln\left(1 + \sqrt{2} \cdot \frac{\lambda}{\xi}\right) \tag{14}$$

Where the logarithmic multiplicative term of $\ln\left(1 + \sqrt{2} \cdot \frac{\lambda}{\xi}\right)$ is the same as that in Eq. 9. The natural appearance of $\ln\left(1 + \sqrt{2} \cdot \frac{\lambda}{\xi}\right)$ in Eq. 12 is a direct consequence of the model, which postulates that the ring magneton has constant current, $I_c(r)$ (Fig. 2,a), independent of its radius. In low-κ superconductors the disk is a very thin hoop, while for high-κ superconductors the disk is a large ring with a tiny central hole. It should be noted that due to the ring geometry, the magnetic flux density, $B$, is concentrated inside the inner diameter of the ring.

The alternative model of a solid magneton can also be considered (Fig. 2,b). In this case, the tangential charge velocity, $v(r)$ is a linear function of $r$ (Fig. 2,b). Thus, to keep the electric current, $dI$, carried by the elemental concentric coil with radius $r$ and width $dr$ constant, the surface charge carrier density should be a reciprocal function of $r$. This model also gives the multiplicative pre-factor of $ln\left(1+\sqrt{2}\cdot\frac{\lambda}{\xi}\right)$ for the Bohr magneton $\frac{\hbar \cdot e}{2 \cdot m_e}$ value. However, the important feature of both models is the ring spatial confinement of the Cooper pair center of mass.

Because the two charge carriers of the Cooper pairs are moving in opposite directions, the net current within a ring is zero, if an external field is not applied. When a field is applied, then the center of mass of the Cooper pairs precesses and this creates the diamagnetic moment. Thus, we propose that the essential condition at which the multiplicative pre-factor of $ln\left(1+\sqrt{2}\cdot\frac{\lambda}{\xi}\right)$ appears in Eqs. 4,8,10 is a new topological state for paired charge carriers (i.e., a ring in 2D case).

## 4. Ring toss model for magnetic flux tube in superconductor

If the applied field at the sample surface, $B_{appl}$, exceeds $B_{c,MO}(T)$, than the magnetic flux penetrates into the superconductor in the form of magnetic flux tubes. In our model, because the Cooper pair center of mass does not occupy a space inside the ring, it is natural to expect that the magnetic flux tube will be trapped there, i.e. magnetic flux will be concentrated inside of a ring within a circle with radius ξ. As far as Cooper pairs rings will be rearranged within the whole sample volume, then the flux tube will appear in the usual manner of a vortex line. Our interpretation of flux lines is similar, but not exactly identical, to the concept of the Abrikosov vortex [21]. The difference is that in the Abrikosov vortex model, the superconducting state inside of the core of radius $\xi$ is completely suppressed. However, the flux density which can suppress the superconducting state should be about the Pauli depairing field, $B_p(T)$:

$$B_p(T) = \frac{2 \cdot \Delta(T)}{g \cdot \mu_B} \gg 4 \cdot B_{c1}(T) = 4 \cdot B_{c,MO} \tag{15}$$

where $\Delta(T)$ is the superconducting energy gap, and $g = 2$, which is at least one order of magnitude greater than the field at the radius $R = \xi$ for the core of the single Abrikosov vortex (see, for instance Eq. 12.18 in [22]):

$$B(r = \xi) = B_{c1} \cdot \frac{K_0\left(\frac{\xi}{\lambda}\right)}{\frac{1}{2} \cdot ln\left(\frac{\lambda}{\xi}\right)} \approx 2 \cdot B_{c1} \cdot \frac{2 \cdot ln\left(\frac{\lambda}{\xi}\right) - 0.6}{ln\left(\frac{\lambda}{\xi}\right)} \lesssim 4 \cdot B_{c1} = 4 \cdot B_{c,MO} \tag{16}$$

where $K_0(x)$ is a zeroth-order modified Bessel function.

In our model, if the applied field $B_{appl}$ is not so high that the single vortex model can be a good approximation, the superconducting state inside of the flux tube is not destroyed by the field, but instead the superconducting state does not exist inside of the tube because of Cooper pair ring spatial confinement. The flux tube, thus, holds surrounding tossed rings of Cooper pairs. It should also be mentioned that Cooper pairs can persistently surround columnar non-superconducting nanowires embedded in the superconducting matrix [23-28].

5. **Critical current density**

Taking into account that:

$$B_{c,MO}(T) = \mu_0 \cdot J_c(T) \cdot \lambda(T) \tag{17}$$

where $J_c(T)$ is the maximum dissipation-free current density, and combining with Eq. 8 one can derive the equation:

$$J_c(T) = \frac{n(T)}{\lambda(T)} \cdot \left[\mu_B \cdot ln\left(1 + \sqrt{2} \cdot \frac{\lambda(T)}{\xi(T)}\right)\right] = \sqrt{\frac{2}{\mu_0 \cdot e^2 \cdot m_e}} \cdot \left[\mu_B \cdot ln\left(1 + \sqrt{2} \cdot \frac{\lambda(T)}{\xi(T)}\right)\right] \cdot n^{1.5}(T). \quad (18)$$

Eq. 18 shows that the critical current density in the material is proportional to the density of Cooper pairs to the power of 1.5. It can be shown that the low-temperature low-κ limit of Eq. 18 has a pair-breaking form:

$$J_c(0) \sim n(0) \cdot \Delta(0) \quad (19)$$

where $\Delta(0)$ is the ground state energy gap. Thus,

$$J_c(T \to 0) = \frac{n(T \to 0)}{\lambda(T \to 0)} \cdot \left[\mu_B \cdot ln\left(1 + \sqrt{2} \cdot \frac{\lambda(T \to 0)}{\xi(T \to 0)}\right)\right] \cong \left[\frac{\pi \cdot e}{\sqrt{2} \cdot m_e}\right] \cdot \frac{1}{v_F} \cdot n(0) \cdot \Delta(0) \quad (20)$$

where, $\xi(0) = \frac{\hbar \cdot v_F}{\pi \cdot \Delta(0)}$ and $v_F$ is the Fermi velocity in the material. Despite that for low-κ materials $J_c$ can be expressed in the form of pair-breaking concept (Eqs. 19,20), the dependence of $J_c$ on the density of Cooper pairs to the power of 1.5 is universal for both low-κ and high-κ materials.

## 6. Discussion

It should be noted that neither Cooper [29] nor Bardeen, Cooper and Schrieffer [4] specified the geometry or spatial confinement for the Cooper pairs. Hirsch [5] proposed a geometrical model for a Cooper pair, where two charge carriers circulate in opposite directions in orbits of radius $2 \cdot \lambda$ with the centers of their orbits separated by ξ. However, this model cannot give the multiplicative pre-factor of $ln\left(1 + \sqrt{2} \cdot \frac{\lambda}{\xi}\right)$ to the Bohr magneton, until the idea of the spatial confinement of the center of mass of the charge carriers within a disc with inner radius ξ and outer radius of $\xi + \sqrt{2} \cdot \lambda$ is implemented.

A change in the system topology at a phase transition is a common feature which has been first proposed by Berezinsky [30,31], and Kosterlitz and Thouless [32]. Taking this into account, our finding, that the superconducting transition is not only associated with the creation of Cooper pairs, but also with the enriching system topology by new objects with genus 1, agrees with general physical principles.

It is interesting to note that our ring model provides a simple geometrical justification for the division between low-κ and high-κ materials. This boundary is set at:

$$\kappa = \left(\frac{1}{2}\right)^{\frac{1}{2}} \quad (21)$$

Thus, at this κ value the ring width, $w$, is equal to the inner radius:

$$w = \left(\xi + \sqrt{2} \cdot \lambda\right) - \xi = \sqrt{2} \cdot \kappa \cdot \xi = \sqrt{2} \cdot \frac{1}{\sqrt{2}} \cdot \xi = \xi \quad (22)$$

This means that the ring in low-κ materials (Type-I) satisfies the geometrical confinement:

$$w \leq \xi \quad (23)$$

while in high-κ (Type-II) materials:

$$\xi \leq w \quad (24)$$

A schematic representation of the ring with $\kappa = \left(\frac{1}{2}\right)^{\frac{1}{2}}$ for which the outer radius:

$$\left(\xi + \sqrt{2} \cdot \lambda\right) = \xi + (2)^{\frac{1}{2}} \cdot \left(\frac{1}{2}\right)^{\frac{1}{2}} \cdot \xi = 2 \cdot \xi \quad (25)$$

and the inner radius is $\xi$ is shown in Fig. 3.

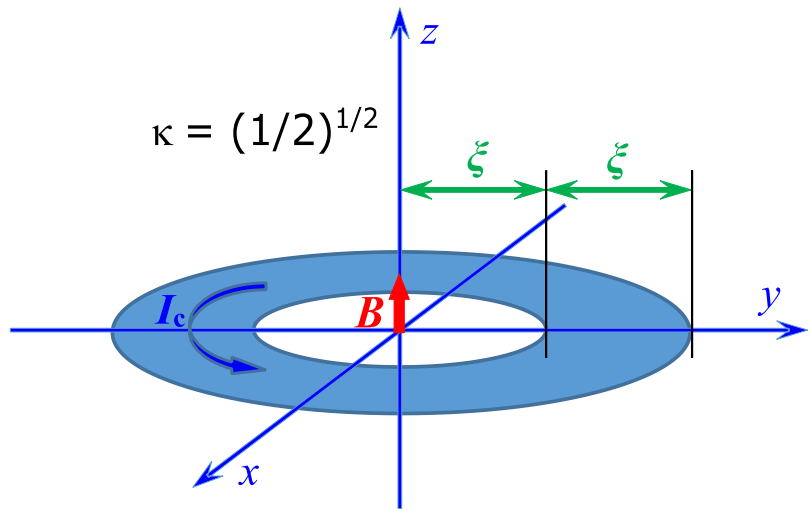

**Figure 3.** Schematic representation of the Cooper pair ring for $\kappa = \left(\frac{1}{2}\right)^{\frac{1}{2}}$. The $I_c$ represents the net current originating from the circulation of the Cooper pair center of mass.

Returning now to the global response of the superconducting sample to an applied magnetic field, $B_{appl}$, we should mention that Condon [33] had proposed an empirical expression for the ground state $B_{c,MO}(0)$ which covers elemental metallic superconductors (i.e., Pb, Ta, Sn, Hg, In, Tl, Nb, and Zr which all are low-κ materials) known at that time:

$$B_{c,MO}(0) = \mu_0 \cdot \mu_B \cdot \delta \cdot \frac{1}{5 \cdot N} \qquad (26)$$

where δ is the number of atoms per unit volume, and $N$ = ½, 1, 2, 3, 4. A comparison of Eq. 8 with Eq. 26 shows that:

$$\delta \cdot \frac{1}{5 \cdot N} = n \cdot ln(1 + \sqrt{2} \cdot \kappa) \qquad (27)$$

where δ and $n$ have similar physical meaning for the density of elemental diamagnets per unit volume, and multiplicative pre-factors of $\frac{1}{5 \cdot N}$ and $ln(1 + \sqrt{2} \cdot \kappa)$ represent the "strength" of the Cooper pair ring in comparison with the classical Bohr magneton.

7. **Summary**

In this paper we note that neither Cooper [29] nor Bardeen, Cooper and Schrieffer [4] specified the topology for Cooper pairs in superconductors. As a result of our analysis, we propose a Cooper pair ring topological state which has spatial confinement of the center of mass within the inner radius ξ and outer radius $\xi + \sqrt{2} \cdot \lambda$. This means that the superconducting transition is associated with the change in system topology by the appearance of Cooper pair rings which exhibit a topological state of genus 1. A universal equation for the Meissner-Ochsenfeld critical field, $B_{c,MO}$, which is based on the concept of the Cooper pair ring and which covers both low-κ and high-κ superconductors has been proposed.

**Acknowledgements**

The author acknowledges financial support provided by the state assignment of Minobrnauki of Russia (theme "Pressure" No. AAAA-A18-118020190104-3) and by Act 211 Government of the Russian Federation, contract No. 02.A03.21.0006.